\documentclass[preprint2]{aastex631}

\submitjournal{ApJ}

\usepackage{lipsum}

\usepackage{amsmath}
\let\fixmathlinenumbering\relax
\usepackage{amssymb}
\usepackage{array}
\usepackage{booktabs}   
\usepackage{url}

\newcommand{\Rp}{R_{\rm p}}

\newcommand{\Rearth}{R_{\oplus}}
\newcommand{\CH}{CH$_4$}
\newcommand{\CO}{CO$_2$}
\newcommand{\Oz}{O$_3$}
\newcommand{\Ot}{O$_2$}

\newcommand{\HzO}{H$_2$O}

\begin{document}

\title{
The First Remotely Detected Biosignature May Not Be the Most Common: Implications for JWST and HWO}

\author{Ravi Kopparapu}
\affiliation{NASA Goddard Space Flight Center, 8800 Greenbelt Rd, Greenbelt, MD, USA}

\begin{abstract}
The first detected member of a new astronomical class is often not
representative of the underlying population, but instead reflects the
selection effects of the observing technique that found it. We apply
this idea to the first remote detection of biosignatures with two
leading near future strategies: JWST transmission spectroscopy and HWO
reflected light direct imaging. Using the known signal scalings of the
two methods together with a simple detectability model, we
show how a rare but observationally favored planet class can dominate
early detections even when it is intrinsically uncommon. For JWST, an
early biosignature detection is most likely to arise from a
detectability favored outlier, such as a sub-Neptune or other
atmosphere rich planet around a nearby M dwarf, rather than from a true
Earth analog. For HWO, the situation is subtler. Among accessible habitable-zone
targets around FGK-type stars, differences in maximum observable
distance and hence in effective survey volume may be smaller than in
the JWST case, weakening the volume bias. At the same time,
stellar-type-dependent photochemistry can alter biosignature
abundances, so the first HWO biosignature may emerge from a balance
between photochemical enhancement and geometric accessibility. Nevertheless, within the accessible sample,
planets with stronger biosignature features and higher reflected light
contrast may still be favored in early detections. A first HWO biosignature could be a selection favored outlier and should  not be assumed to represent inhabited rocky planets in general. Crucially, the
longest lived biosphere on a planet is not necessarily its most
spectrally detectable one. If the first detection turns out to be an
outlier, that may still suggest that a more broader range of
habitable environments awaits discovery.
\end{abstract}



\section{Introduction} \label{sec:intro}
New observational capabilities often uncover not a representative sample of the universe from their first detection, but its most detectable extremes. This follows from sensitivity limited surveys: when objects span a range of signal strengths, the brightest, strongest, or otherwise most observable members are found first, while the more common but weaker members remain hidden. This selection effect was formalized long ago as Malmquist bias \citep{malmquist1922}. Some examples from the history of first detections in astronomy illustrate this pattern repeatedly:
\begin{itemize}

\item Scorpius X-1 (Sco-X-1) was the first X-ray source discovered outside the solar system \citep{xray1962}, and is the brightest non-solar X-ray source in sky.
\item 3C 273 is the first quasar  to be discovered \citep{q1,q2}, and is the brightest optical quasar seen from the Earth.
\item The first radio source outside the solar system was detected \citep{jansky1933} from the center of the Milky Way, even though currently it is not the brightest object in the sky. At low frequencies that Jansky used, Cygnus A is the brightest source. However, Jansky's instrument had a lower resolution and wider beam-width making more diffused emissions from large scale structures like Milkyway to be easily detectable rather than point sources like Cygnus A.  
\item The first fast radio burst (FRB) (``Lorimer Burst'', \cite{lorimer2007}) is an extreme outlier among the early sources detected by the Parkes radio observatory. It was found in the archival data with a burst duration of $< 5$ms, and was remarkably bright for its time of discovery. 
\item Ceres is the first asteroid/dwarf-planet discovered by Giuseppe Piazzi in 1801, and is the largest asteroid. 
\item The first confirmed exoplanet orbited a millisecond pulsar \citep{wolszczan1992}. Neither pulsars, nor the pulsar planets are the most common type of objects in their category. However, the combination of pulsar planets produce the brightest (``loudest'') signal with the corresponding sensitive instrument (a radio telescope that can detect deviations in timed pulsed emissions). 
\item The first hot Jupiter \citep{mayor1995}, 51~Pegasi~b, was a gas giant on a 4 day orbit, a geometric extreme that maximized radial velocity amplitude.  Hot Jupiters represent $\sim\! 0.5 - 1\%$ of planetary systems \citep{wright2012, dawson2018}, yet they dominated early exoplanet catalogs. In fact, the existence of hot-jupiters were actually predicted by \cite{struve1952}, mentioning that ``It is not unreasonable that a planet might exist at a distance of 1/50 astronomical unit...period around a star of solar mass would then be about 1 day''.
\end{itemize}

\footnote{The same logic may even apply to stars themselves. The first stellar object available to detailed observation was the Sun, the brightest and most accessible star from Earth, even though it is not representative of the broader stellar population.} In each case, the first detected member of a new class was the \emph{loudest} one accessible to the instrument at hand, rather than the most common or representative member of the underlying population. We propose that the first remote biosignature detection on an exoplanet will arise from the same observational selection effect.

This paper addresses a focused and tractable question: Given the two primary near future biosignature search strategies, JWST transmission spectroscopy and HWO reflected light spectroscopy, what biosignature scenario is most detectable for each technique, and how unrepresentative is it likely to be of the broader population of inhabited worlds? The answer can be approached from known physics, instrument capabilities, and atmospheric chemistry alone, without requiring prior knowledge of how common life is.

 Here, the argument concerns the first biosignature signal strong enough to support a serious or widely accepted biological interpretation, not merely the first unusual atmospheric feature.

\citet{kipping2025} recently applied a related ``loudest first'' argument to technosignatures. Here we consider the corresponding problem for remote \emph{biosignatures}, which involves a different observational domain and a different physical basis for detectability. We develop an instrument specific framework in which detectability is decomposed into the underlying geometric, atmospheric, and spectroscopic terms relevant to each observing strategy. In particular, we treat transmission spectroscopy and direct imaging separately, since they are governed by fundamentally different selection effects and therefore different notions of observational loudness. We further examine what a first HWO detection would imply, regardless of which Earth-like atmospheric state is ultimately observed.

\section{Detection Methods and Signal Scalings}
\label{subsec:detection_scalings}

In this section we summarize the signal-to-noise scalings
for the two observing strategies considered in this paper:
transmission spectroscopy with JWST-like facilities and reflected-light
direct imaging with HWO-like facilities. These relations are not new;
they are included here only to establish the different observational
selection functions of the two techniques before turning, in the next
section, to the question of which atmospheres are most likely to be
detected first.

\subsection{Transmission spectroscopy}
\label{subsec:transmission_scaling}

For transmission spectroscopy, the wavelength-dependent excess transit
depth of an atmospheric spectral feature can be approximated as
\begin{equation}
  \Delta \delta_\lambda
  \;\approx\;
  \frac{2 R_p z_\lambda}{R_\star^2},
  \label{eq:transmission_signal}
\end{equation}
where $R_p$ is the solid body planet radius, $R_\star$ is the stellar radius, and
$z_\lambda$ is the effective atmospheric height at wavelength
$\lambda$. To leading order, the atmospheric scale height is
\begin{equation}
  H \;=\; \frac{k_B T}{\mu g},
  \label{eq:scale_height}
\end{equation}
with $T$ the atmospheric temperature, $\mu$ the mean molecular mass,
and $g$ the surface gravity. For a strong spectral band,
$z_\lambda$ is typically of order a few scale heights,
\begin{equation}
  z_\lambda \sim N_H H,
  \label{eq:effective_height}
\end{equation}
where $N_H$ is an order-unity factor that depends on the opacity
structure, abundance of the absorber, and cloud or haze opacity.

If the fractional uncertainty in the measured spectral transit depth
per transit is $\sigma_{\rm tr}$, then after $N_{\rm tr}$ observed
transits the signal-to-noise ratio scales approximately as
\begin{equation}
  {\rm SNR}_{\rm tr}
  \;\approx\;
  \frac{\Delta \delta_\lambda}{\sigma_{\rm tr}}
  \sqrt{N_{\rm tr}}.
  \label{eq:snrtr}
\end{equation}
Here $\sigma_{\rm tr}$ has the same units as $\Delta\delta_\lambda$;
that is, it is the fractional uncertainty on the measured excess
transit depth in the relevant spectral channel.
In the photon limited approximation, the per transit uncertainty
$\sigma_{\rm tr}$ is set primarily by photon noise from the host star.
In that limit,
$\sigma_{\rm tr} \propto F_\star^{-1/2}$, where $F_\star$ is the
stellar flux received at Earth. Since $F_\star \propto L_\star D^{-2}$,
with $L_\star$ the stellar luminosity and $D$ the distance to the
system, it follows that
\begin{equation}
  \sigma_{\rm tr} \;\propto\; \frac{D}{L_\star^{1/2}}.
  \label{eq:sigma_tr}
\end{equation}
This has two immediate consequences. First, nearer and intrinsically
brighter host stars yield lower $\sigma_{\rm tr}$ and hence higher
SNR for a given atmospheric signal. Second, substituting
Equation~(\ref{eq:sigma_tr}) into Equation~(4), the transmission
spectroscopy SNR scales as
\begin{equation}
  {\rm SNR}_{\rm tr}
  \;\propto\;
  \frac{\Delta\delta_\lambda \, L_\star^{1/2} \, N_{\rm tr}^{1/2}}{D}.
  \label{eq:snr_explicit}
\end{equation}
This makes explicit that transmission spectroscopy favors not only
planets with large spectral feature amplitudes $\Delta\delta_\lambda$
and systems with many observable transits $N_{\rm tr}$, but also bright,
nearby host stars. For M dwarfs, the low stellar luminosity partly
reduces this advantage; however, for nearby systems this effect is often
outweighed by the much stronger geometric gain from the $R_\star^{-2}$
dependence in Equation~(1). In practice, $\sigma_{\rm tr}$ can also
include contributions from instrumental systematics and stellar
variability, but the photon limited scaling above is sufficient for the
leading-order argument developed here.

These expressions show the well-known result that transmission
spectroscopy strongly favors small host stars, atmospheres with large
effective scale heights, and systems for which multiple transits can be
observed over the mission lifetime.

Metrics of this general kind have already been used in the literature
to rank targets for atmospheric characterization. For example,
\citet{kempton2018} introduced the Transmission Spectroscopy Metric
(TSM) as an analytic proxy for the relative suitability of planets for
transmission spectroscopy, based on the same leading order dependence on
planetary size, atmospheric scale height, stellar radius, and host star
brightness. Our goal here is different: rather than ranking targets for
follow-up in general, we use the same underlying detectability logic to
ask what kinds of atmospheres are most likely to dominate the first
biosignature detections.

\subsection{Reflected-light direct imaging}
\label{subsec:direct_scaling}

For reflected-light direct imaging, the fundamental observable is the
planet--star flux contrast \citep[Appendix, Eq.(11)]{robinson2016},
\begin{equation}
  C_p(\lambda)
  \;=\;
  A_g(\lambda)\,\Phi(\alpha)
  \left(\frac{R_p}{a}\right)^2,
  \label{eq:contrast_signal}
\end{equation}
where $A_g(\lambda)$ is the geometric albedo, $\Phi(\alpha)$ is the
phase function at phase angle $\alpha$, and $a$ is the orbital
separation.

A biosignature absorption feature is then described as a fractional
depression of this reflected light continuum. If the feature has
fractional depth $A_{\rm bio}(\lambda)$ relative to the local
continuum, its approximate amplitude in planet star contrast is

\begin{equation}
  \Delta C_{\rm bio}(\lambda)
  \;\approx\;
  C_p(\lambda)\,A_{\rm bio}(\lambda).
  \label{eq:contrast_feature}
\end{equation}

Direct detection additionally requires that the planet be angularly
separated from the host star:

\begin{equation}
\theta \gtrsim {\rm IWA} \approx N_{\rm IWA}\frac{\lambda}{D_{\rm tel}},
\label{eq:iwa}
\end{equation}
where $D_{\rm tel}$ is the telescope diameter, IWA is the inner working angle of the instrument and $N_{\rm IWA}$ is a
design dependent factor of order a few.


If the effective uncertainty in the spectral channel is
$\sigma_{\rm dir}$, then after integration time $t_{\rm obs}$ the
signal-to-noise ratio scales approximately as

\begin{equation}
{\rm SNR}_{\rm dir}=
\begin{cases}
\dfrac{\Delta C_{\rm bio}}{\sigma_{\rm dir}}\sqrt{t_{\rm obs}},
& \theta \ge {\rm IWA},\\[10pt]
0, & \theta < {\rm IWA}.
\end{cases}
\label{eq:direct_snr_piecewise}
\end{equation}

In this simplified treatment, we approximate the coronagraphic accessibility as a hard cutoff at the inner working angle: planets interior to the IWA are taken to be undetectable, while planets exterior to the IWA are treated with the leading order signal scaling above.
These relations show the familiar result that direct imaging is
governed primarily by reflected light contrast, biosignature band
depth, and angular separation. 

Unlike transmission spectroscopy, direct imaging is not helped by small
stellar radius in the same direct geometric way. Instead, detectability
is governed primarily by reflected-light contrast, spectral feature
depth, and angular separation, so habitable zone planets around late
M dwarfs are often disfavored because their separations fall inside the
coronagraph inner working angle.


\subsection{A Simple First-Detection Bias Model}
\label{subsec:toy_model}
The main idea of this paper is that the first detected
biosignature is likely to come from a planet that is observationally
favorable rather than from a planet that is more
common.  In this subsection we make that argument quantitative using
a simple two class model.
 
 
Let inhabited planets belong to two broad classes.  The first
class, denoted $q$ (quiet), is intrinsically common but less
detectable.
The second class, denoted $l$ (loud), is intrinsically rarer but more
detectable.
Let $n_q$ and $n_l$ be the number densities of each class
(occurrence rate times host star density\footnote{This can be
interpreted as the product of the first three terms in the Drake
equation.}).
 
For a given observing strategy, define $\rho_j(d)$ as the detection
statistic (likely SNR, as given in Eq.(\ref{eq:snr_explicit}) ) that a planet of
class $j$ produces at distance $d$.  For a survey with a fixed
detection threshold $\rho_{\rm thresh}$, a planet is detectable only
when $\rho_j(d) \geq \rho_{\rm thresh}$.  The \emph{effective survey
volume} for class $j$ is the spherical volume of space within which its members
are detectable:
\begin{equation}
  V_{{\rm eff},j}
  \;=\;
  \frac{4\pi}{3}\,d_{{\rm max},j}^{\,3},
  \label{eq:veff}
\end{equation}
where $d_{{\rm max},j}$ is the maximum distance at which a class $j$
planet lies above the detection threshold.  The expected number of
detections from each class then scales as
\begin{equation}
  N_{{\rm det},j} \;\propto\; n_j\,V_{{\rm eff},j},
  \label{eq:ndet}
\end{equation}
and the ratio of detections between the two classes is
\begin{equation}
  \frac{N_{{\rm det},l}}{N_{{\rm det},q}}
  \;=\;
  \frac{n_l}{n_q}
  \left(\frac{d_{{\rm max},l}}{d_{{\rm max},q}}\right)^{\!3}.
  \label{eq:ratio}
\end{equation}
This ratio is the key result. To see why loud planets dominate, note
that a louder planet, one with higher intrinsic detectability (or SNR)$\rho_0$, produces a
higher SNR at any given distance, and therefore remains above the
detection threshold $\rho_{\rm thresh}$ out to a greater maximum
distance $d_{\rm max}$. Since the effective survey volume grows as
$d_{\rm max}^3$ (Equation~\ref{eq:veff}), a class with larger $\rho_0$
is detectable within a proportionally larger volume of space. The loud
class therefore sweeps a larger portion of the sky even if its members
are individually rarer. The first detections are biased towards the class that maximizes the expected number of detections, which is often the rarer but louder case.

 Eq.(\ref{eq:ratio}) can be rewritten in terms of the intrinsic detectability

\begin{equation}
  \frac{N_{{\rm det},l}}{N_{{\rm det},q}}
  \;=\;
  \frac{n_l}{n_q}
  \left(\frac{\rho_{0,l}}{\rho_{0,q}}\right)^{\!3}.
  \label{eq:ratio_final}
\end{equation}

It says: the loud class
dominates first detections whenever
$n_l/n_q > (\rho_{0,q}/\rho_{0,l})^3$.  Or equivalently, the loud
class can afford to be rarer than the quiet class by up to a factor of
$(\rho_{0,l}/\rho_{0,q})^3$ and still be equally likely to be detected
first.

We note an important caveat in applying Equation~(\ref{eq:ratio_final})
to direct imaging with HWO. For transmission spectroscopy, the SNR scales as $D^{-1}$
(Equation~6). Let $\rho_0$ be the detectbility 
(SNR) at a reference distance $D_0$. Then
\begin{equation}
  \mathrm{SNR}(D) = \rho_0 \frac{D_0}{D}.
\end{equation}
Setting this equal to the detection threshold
$\rho_\mathrm{thresh}$ gives the maximum detectable distance,
\begin{equation}
\begin{split}
  d_\mathrm{max}
  &= D_0 \frac{\rho_0}{\rho_\mathrm{thresh}} \\
  &\propto \rho_0 .
\end{split}
\end{equation}
The effective survey volume therefore scales as
\begin{equation}
  V_\mathrm{eff} \propto d_\mathrm{max}^3 \propto \rho_0^3.
\end{equation}

 To illustrate the magnitude of this effect, consider a K2-18b-like
sub-Neptune ($R_p \approx 2.6\,R_\oplus$, H$_2$-rich atmosphere with
mean molecular weight $\mu \approx 2.3$) versus a rocky Earth analog
($R_p = 1.0\,R_\oplus$, N$_2$-dominated atmosphere with
$\mu \approx 28$), both orbiting a similar nearby M dwarf. From
Equations~(1)--(3), the spectral feature amplitude scales as
$\Delta\delta_\lambda \propto R_p z_\lambda/R_\star^2$, where
$z_\lambda \sim N_H H$. In the simplified case where temperature,
gravity, and host star properties are similar\footnote{Note that the gravity `g' is not similar, because the gravity of K2-18b is $\sim 1.32$ times Earth gravity. If we include  $g$ as well in $z_{\lambda}$, then $z_{\lambda} \propto \mu^{-1} g^{-1}$, the loudness ratio $\frac{\rho_{0,l}}{\rho_{0,q}}$ in the following equations becomes $\approx 24$, still similar and within the range of 32 derived below.}, this
gives $z_\lambda \propto \mu^{-1}$.

Because the loudness, $\rho$, is essentially SNR as mentioned above in the beginning of this sub-section, we can use Eq.(\ref{eq:snr_explicit}) assuming that $L_{\star}$, $N_\mathrm{tr}$ and $D$ are same for both K2-18b and an Earth-analog.

Then, the leading order loudness
ratio  becomes
\begin{equation}
\begin{aligned}
\frac{{\rm SNR}_{l}}{{\rm SNR}_{q}}
&\;\approx\;
\frac{\rho_{0,l}}{\rho_{0,q}} \\
&\;\approx\;
\frac{R_{p,l}}{R_{p,q}} \times \frac{\mu_q}{\mu_l}
\;=\; \frac{2.6}{1.0} \times \frac{28}{2.3}
\;\approx\; 32.
\end{aligned}
\label{eq:loudness_ratio}
\end{equation}
Substituting into Equation~(\ref{eq:ratio_final}) gives
\begin{equation}
\frac{N_{\rm det,l}}{N_{\rm det,q}}
\;\approx\;
\frac{n_l}{n_q} \times 32^3
\;\approx\;
\frac{n_l}{n_q} \times 3\times10^4.
\label{eq:worked_example}
\end{equation}
The two classes contribute equally to the expected early detection
yield when $N_{\rm det,l}/N_{\rm det,q} = 1$, which requires the ratio of the number densities of such kind planets to be
\begin{equation}
\frac{n_l}{n_q} = \frac{1}{32^3} \approx \frac{1}{3\times10^4}.
\label{eq:breakeven}
\end{equation}
In other words, the sub-Neptune class (`loud', $l$) and the rocky Earth-analog
class (`quiet', q) become equally competitive in the early-detection statistics
when inhabited sub-Neptunes are $\sim\!3\times10^4$ times rarer than
inhabited Earth analogs. If inhabited sub-Neptunes are more common
than this break-even ratio—even if still much rarer in absolute
terms—they are more likely to dominate the earliest detections.
Conversely, only if they are rarer than about one in
$3\times10^4$ relative to Earth analogs would the quieter rocky-planet
class be favored. The key point is that even a moderate intrinsic
loudness advantage, once amplified by the cubic survey volume scaling,
can overcome very large differences in occurrence rate.

For direct imaging, the situation is more
complex. As a system is moved to greater distance, three effects act
simultaneously: the planet flux decreases as $D^{-2}$, the angular
separation $\theta=a/D$ shrinks toward the IWA, and exozodiacal and
background noise terms become increasingly important relative to the
planet signal. If the IWA constraint is the dominant factor, as may
well be the case for habitable zone rocky planets around nearby
Sun-like stars, whose angular separations are already close to HWO's
expected IWA, then $d_{\rm max}$ is set primarily by orbital geometry
and telescope design rather than by the atmospheric loudness $\rho_0$.
In that limit, planets in similar orbits have comparable
$d_{\rm max}$, the ratio $d_{\rm max,l}/d_{\rm max,q}\approx 1$, and
Equation~(\ref{eq:ratio_final}) reduces to
$N_{\rm det,l}/N_{\rm det,q}\approx n_l/n_q$: the most common inhabited
atmospheric state would be found first, not necessarily the loudest
one.

The degree to which the IWA constraint or the SNR noise floor sets
$d_{\rm max}$ for a given HWO survey depends on the specific telescope
aperture, coronagraph design, target distances, and planetary system
architectures, and a full treatment is beyond the scope of this paper.
The truth is likely intermediate: both effects contribute, and their
relative importance varies across the survey sample. Even in an
IWA-dominated regime, however, a residual selection effect remains. The
IWA determines which planets are accessible at all, but among those
accessible planets the reflected light SNR still depends on quantities
such as the biosignature feature depth $A_{\rm bio}$
(Equation~\ref{eq:contrast_feature}), the geometric albedo $A_g$
(Equation~\ref{eq:contrast_signal}), and the phase function
$\Phi(\alpha)$. Within the IWA-accessible sample, planets with stronger
biosignature features can still reach detectable SNR in shorter
integration times and may therefore appear earlier in a time-ordered
survey. The selection effect for HWO is thus not the same strong
$\rho_0^3$ scaling as in the JWST transit case, but it does not vanish
entirely. The first HWO biosignature detection is still expected to
favor the most observationally accessible atmospheric state within the
survey sample, even if the geometric selection is weaker than in the
transit case.

It is worth noting that the longevity of a biosphere state on a planet
does not by itself determine whether it will be the first detected.
On Earth, the Archean era, characterized by an anoxic atmosphere and
high biogenic methane production, lasted for roughly for 1.5 billion years (4.0--2.5 Gyr)),
making it the longest inhabited phase in Earth's history. Yet that does
not necessarily make it the most detectable phase for a telescope like
HWO. A longer lived biosphere state contributes a larger prior
probability of being observed at a random epoch, but it does not by
itself increase the spectral contrast of its biosignature features.
Longevity and detectability are therefore distinct quantities. A planet
could host a relatively short lived but spectrally prominent biosphere
that is easier to detect than a longer lived but spectrally quieter one.
For HWO like observations, detectability will depend on how strongly a
given atmospheric state expresses its biosignature features across the
mission's wavelength range, as well as on albedo, cloud properties, and
instrument sensitivity. 

\begin{figure*}
\centering
\includegraphics[width=\textwidth]{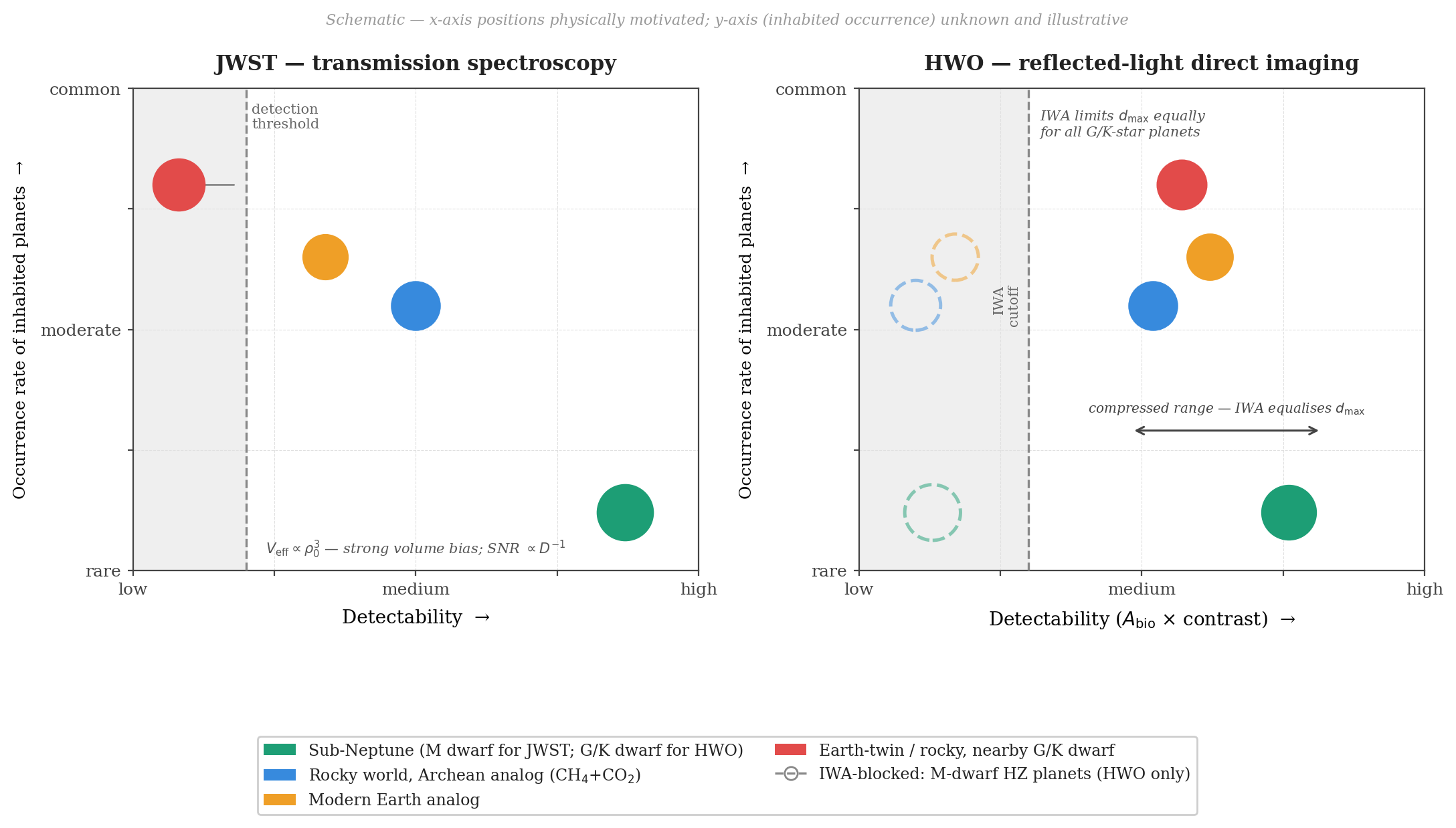}
\caption{Schematic illustration of occurrence rate of inhabited planets versus
observational detectability for JWST transmission spectroscopy
(\textit{left}) and HWO reflected-light direct imaging
(\textit{right}). Circle positions on the $x$-axis are physically
motivated by the signal scalings of Section~\ref{subsec:detection_scalings};
$y$-axis positions are illustrative: the true occurrence rate of
inhabited planets in any class is unknown. \textit{Left:} The strong
$V_\mathrm{eff} \propto \rho_0^3$ volume bias
(Equations~\ref{eq:veff}--\ref{eq:ratio_final}) places the
sub-Neptune class (green) near the top of the JWST detectability
hierarchy despite its assumed rarity as an inhabited world.
Earth twin planets around Sun-like stars (red) fall below the
detection threshold entirely, regardless of their occurrence rate.
\textit{Right:}  habitable zone planets that are further away (dashed circles,
shaded region) are blocked by HWO's coronagraph inner working angle
($\theta \lesssim IWA $), including
sub-Neptunes. Among accessible G/K star targets at ${\sim}1$~AU,
the IWA can limit $d_\mathrm{max}$ similarly across all planet types,
compressing the detectability range relative to the JWST panel and
weakening the $\rho_0^3$ volume bias. The residual spread reflects
differences in biosignature feature depth $A_\mathrm{bio}$
(Equation~\ref{eq:contrast_feature}) and planet/star contrast.  In both
panels, the first biosignature detection is expected to be drawn from the high detectability or high accessibility tail, not from the most common inhabited
world.}
\label{fig:occurrence_detectability}
\end{figure*}

Figure~\ref{fig:occurrence_detectability} illustrates these two
regimes schematically.

\section{FIRST TARGETS FAVORED BY DETECTABILITY}
\subsection{JWST favored First targets}
We suggest that K2-18b, or more broadly this class of planet, may occupy in biosignature science a role analogous to that of 51~Pegasi~b in exoplanet science: the most favorable target for the current instrument generation, extreme along several axes of detectability, and therefore unlikely to be representative of the typical habitable planet, if it does turn out to be habitable.
 
K2-18b \citep{benneke2019,tsiaras2019} is a sub-Neptune
($\Rp \approx 2.6\,\Rearth$, $M \approx 8.6\,M_\oplus$) orbiting an
M2.5 dwarf at $\sim\!0.14$~AU ($P \approx 33$~days) and lying at a
distance of $\sim\!124$~ly. It is a particularly favorable target under
Equation~(\ref{eq:snrtr}): it combines a large planetary
radius, a small host star, a likely H$_2$-rich atmosphere with a large
scale height, and a relatively bright star. Transmission
spectroscopy with JWST has robustly detected \CH{} at $\geq 4\sigma$
and has also identified \CO{} in its atmosphere
\citep{madhusudhan2023}, consistent with a hydrogen-rich atmosphere.

At the same time, K2-18b's status as a biosignature candidate remains
actively debated. \citet{madhusudhan2023} proposed a ``Hycean world''
interpretation, in which a liquid ocean lies beneath an H$_2$-rich
atmosphere, potentially allowing conditions
suitable for microbial life. Other studies, however, have argued that
the observed \CH{} and \CO{} abundances may also be consistent with a
gas-rich mini-Neptune atmosphere that does not require an active
biosphere \citep{wogan2024}; see also \citet{stevenson2025} for a
review. The current observational picture is therefore more limited:
\CH{} is robustly detected, \CO{} is likely present, but there is as
yet no statistically convincing evidence for sulfur-bearing
biosignature molecules. The possibility of a biogenic DMS/DMDS signal
has not been ruled out, but neither has it been established. Further
observations will be required.

Regardless of how the K2-18b interpretation is ultimately resolved, the
system serves an important illustrative role in the present argument. It
is precisely the kind of planet that is expected to produce one of the
first claimed biosignature detections, not because it is necessarily
typical of inhabited worlds, but because it lies near the top of the
current detectability hierarchy for transmission spectroscopy.

 The selection principle illustrated by K2-18b applies more broadly
to planets that score highly under the detectability scaling of
Equation~(\ref{eq:snr_explicit}). Several other current JWST targets occupy similarly
favorable positions in the transmission spectroscopy hierarchy.
LHS~1140~b is a rocky super-Earth ($R_p \approx 1.7\,R_\oplus$)
orbiting a nearby M~dwarf at $\sim$15~pc in the habitable zone; its
small host star and favorable transit depth make it one of the best
characterized rocky habitable zone targets available
\citep{Damiano2024}. The TRAPPIST-1 system provides seven transiting
rocky planets around an ultracool M~dwarf at 12~pc, with habitable zone
members e, f, and g offering repeated transit opportunities that build
signal as $N_{\rm tr}^{1/2}$ over a mission lifetime \citep{gillon2017}.
If biosignatures were detected in any of these systems, they should not
be assumed to represent the broader inhabited-planet population. Like
K2-18b, they are compelling first search targets because they are among
the loudest accessible systems for current instruments, not because
they are necessarily the most representative ones.

\subsection{HWO atmospheric states}
For HWO, ``Earth-like'' does not correspond to a single spectral state.
Earth's atmosphere has evolved substantially over its
$\sim\!4.5$~Gyr history, and different phases of that evolution would
present different remotely detectable biosignatures
\citep{meadows2018,schwieterman2018,schwieterman2024}. The character of a first HWO
biosignature detection therefore depends critically on which
Earth-through-time atmospheric state the first detected rocky planet
happens to exhibit. Just as importantly, these states are not equally
detectable. Recent work shows that broad wavelength coverage from
$0.26$ to $1.7\,\mu$m is needed to identify and properly contextualize
biosignatures across this range of atmospheric histories
\citep{krissansen2025}. Archean-like \CH{}+\CO{} features are expressed
primarily longward of $0.8\,\mu$m, Proterozoic \Oz{} is most readily
identified in the UV, and modern-Earth \Ot{} at $0.76\,\mu$m requires
optical SNR $\gtrsim 20$ \citep{latouf2025,ulses2025}.  The Proterozoic state also presents a false-negative risk: without
UV wavelength coverage to detect the O$_3$ Hartley band, a
low oxygen Proterozoic world could be mistaken for a lifeless
planet \citep{reinhard2017}. The
detectability of a first HWO biosignature therefore depends not only on
the instrument configuration, but also on which atmospheric state is
present.

\begin{figure*}
\centering
\includegraphics[width=\columnwidth]{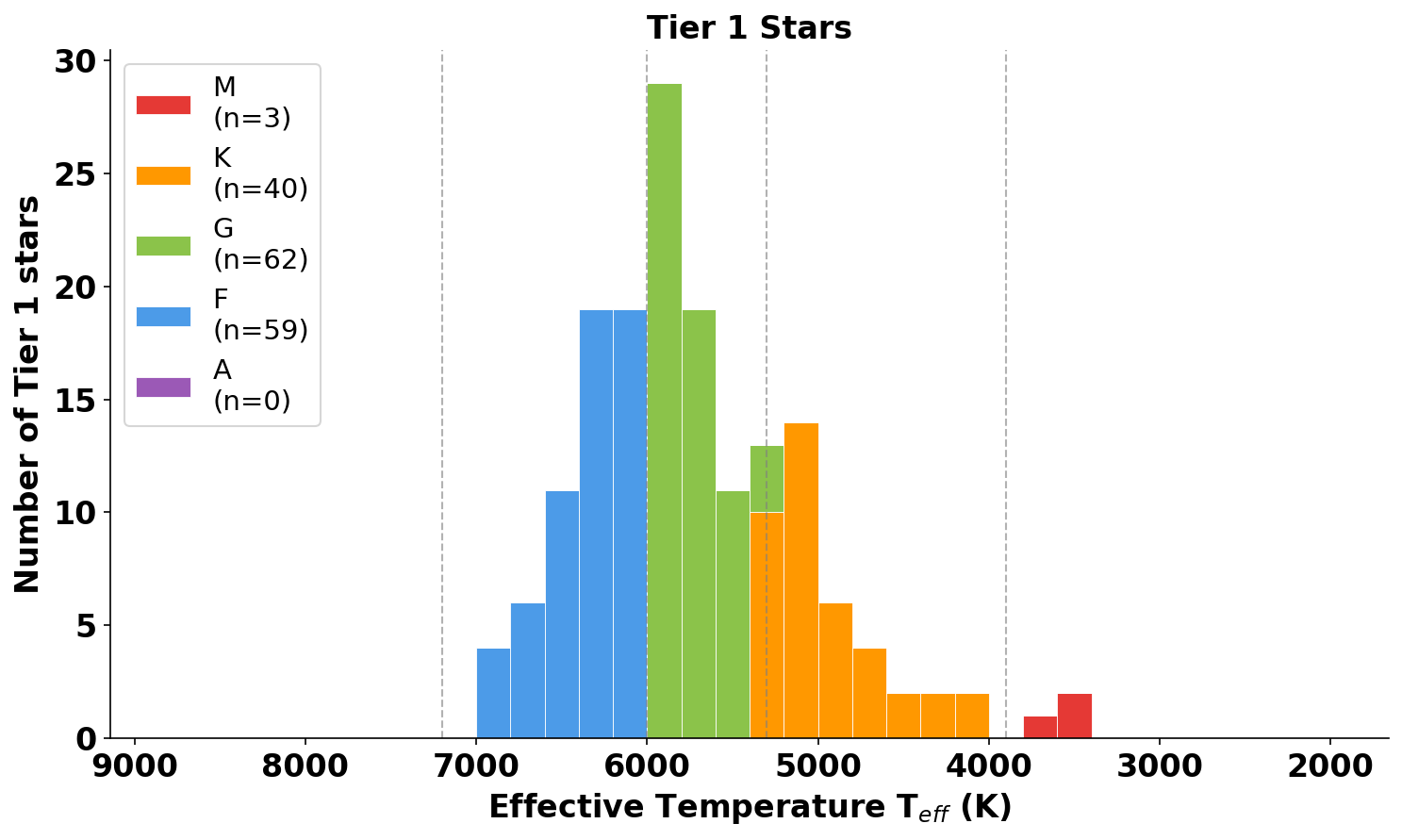}
\includegraphics[width=\columnwidth]{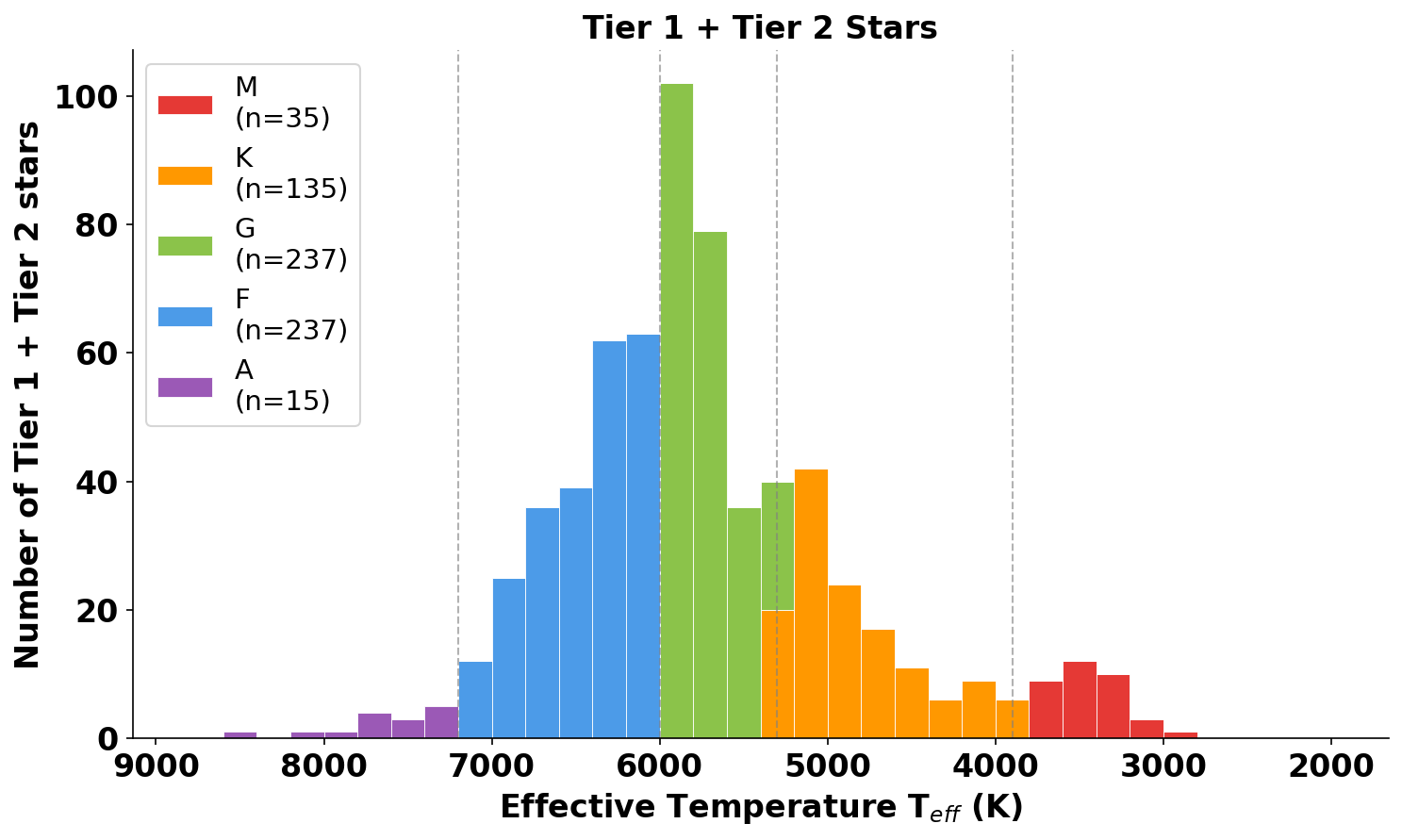}
\caption{ Distribution of effective temperatures of HWO target
stars from the TSS25 catalog \citep{tuchow2025}, color coded by
spectral type, for two priority subsamples. \emph{Left:} Tier~1
stars ($n = 164$), the highest priority sample identified as having the most accessible habitable zones with the lowest exposure times, hence
a high probability of being observed by HWO regardless of the
mission's final architecture \citep[section 2.1]{tuchow2025}. \emph{Right:}
Tier~1 + Tier~2 stars ($n = 644$), which together constitute the
full high probability sample. Tier~2 extends the target list by
incorporating stars that appear in multiple independent yield
calculations across a range of HWO design concepts
\citep[section 2.2]{tuchow2025}. In both panels, F dwarfs
($6000$--$7200$\,K; blue) and G dwarfs ($5300$--$6000$\,K; green)
dominate the samples, while K dwarfs ($5300$--$3900$\,K; orange) and M dwarfs ($3900$--$2300$\,K; blue) come next, respectively, due to fainter magnitudes. 
The near-equal split
between F and G hosts has implications for biosignature
detectability: F and G stars present opposite photochemical
trade offs for biosignature gas accumulation (see text), meaning
the stellar type of the first HWO biosignature host may itself
reflect an observational selection effect.}
\label{fig:hwo_targets}
\end{figure*}

 Figure~\ref{fig:hwo_targets} shows the distribution of effective
temperatures in the TSS25 HWO target star sample from \cite{tuchow2025} which included target star list from \citet{ms2024}.
The accessible target list is weighted toward F and G dwarfs\footnote{ Temperature ranges  for FGKM spectral types were selected from here \url{https://github.com/emamajek/SpectralType/blob/master/EEM_dwarf_UBVIJHK_colors_Teff.txt}} in
roughly equal numbers ($n=59$ and $n=62$, respectively for Tier 1), with K
dwarfs ($n=40$ for Tier 1, and $n=135$ for Tier 2) and M dwarfs ($n=3$  and $n=35$) less represented. This pattern
likely reflects the fact that, although F stars are intrinsically
rarer than G or K stars in the solar neighborhood, their brighter and
more widely separated habitable zones make a larger fraction of their
planets accessible outside the IWA. The near equal representation of
F and G stars has direct consequences for biosignature detectability
through photochemistry. Cooler hosts, especially K dwarfs, can sustain
longer photochemical lifetimes and higher steady-state abundances of
trace gases such as \CH{} for a given biological flux
\citep{segura2005,arney2019}, potentially boosting $A_{\rm bio}$ and
$\rho_0$. However, K-dwarf habitable zones lie at smaller angular
separations and are more vulnerable to the IWA cutoff, reducing
$d_{\rm max}$. F-type stars present the opposite trade-off: their
habitable zones are wider and more accessible, but their higher UV
output tends to shorten the lifetimes of some biosignature gases. The
first HWO biosignature may therefore emerge from a balance between
photochemical enhancement and geometric accessibility across the
accessible FGK target population, and the stellar type of the first
HWO host may itself reflect a selection effect rather than the most
common environment for inhabited planets.

\begin{table*}
\centering
\caption{ Illustrative literature-based comparison of Earth-through-time
biosignature favorability for an HWO-like telescope (6\,m aperture,
$0.2$--$2.0\,\mu$m, $R=7$ UV / $R=140$ VIS / $R=70$ NIR, SNR\,$=$\,20
baseline). The entries are semi quantitative and intended to summarize
broad trends: different atmospheric states place their most informative
biosignature features in different wavelength regions, and HWO
detectability depends on how those features align with the instrument
bandpass and required SNR. The three eras appear  competitive
through different spectral windows rather than one being overwhelmingly
favored. See \citet{arney2016}, \citet{schwieterman2018},\citet{young2024},
\citet{krissansen-totton2025}, \citet{latouf2025}, and
\citet{ulses2025}.}
\label{tab:earthtime}
\setlength{\tabcolsep}{4pt}
\begin{tabular}{lllll}
\hline\hline
Atmospheric  & Biosignature(s) & Wavelength  & HWO  & Comments \\
state&&region&favorability&\\
\hline
Archean & CH$_4$+CO$_2$, &
NIR ($>$0.8\,$\mu$m) &
Moderate/high &
Requires \\
(4.0--2.5 Gyr)& haze, H$_2$O &&&NIR $\gtrsim$1.6\,$\mu$m \\
\hline
Proterozoic &O$_3$, weak O$_2$,  &
UV/optical  &
Moderate &
O$_3$ UV detectable  \\
(2.5--0.541 Gyr)&H$_2$O&(0.2--0.5\,$\mu$m)&& at low O$_2$ \\
\hline
Modern-Earth &
O$_3$, O$_2$, H$_2$O &
UV/optical  &
Moderate/high &
O$_2$/O$_3$ in  \\
(0.541 Gyr--now)&&(0.2--0.8\,$\mu$m)&& favorable windows \\
\hline
\end{tabular}
\tablecomments{ This table is intentionally schematic. The three atmospheric regimes
appear broadly competitive in different parts of the HWO bandpass,
consistent with the compressed detectability range discussed in
Section~2.3 and Figure~1. No single Earth through time state emerges
here as overwhelmingly favored by intrinsic feature strength alone;
rather, detectability depends on a combination of spectral band depth,
wavelength placement, reflected light contrast, and geometric
accessibility. The first HWO biosignature may therefore come from any of these
atmospheric states, but its discovery would not by itself show that
the detected state is the dominant or most representative inhabited
outcome among rocky planets.}
\end{table*}

 Table~\ref{tab:earthtime} summarizes this comparison qualitatively.
Taken at face value, the literature suggests that Archean-,
Proterozoic-, and modern-Earth-like atmospheres may all be broadly
competitive HWO targets, though through different spectral windows and
with different contextual requirements.

In the language of Eq. (\ref{eq:ratio_final}), longevity would enter through the
abundance term ($n$), whereas detectability enters through the characteristic
signal strength ($\rho_{0}$). A longer-lived atmospheric state may therefore still
fail to dominate the earliest detections if its biosignature features
are weaker or less favorably placed within the instrument bandpass than
those of a shorter-lived but more spectrally prominent state. A full
quantitative ranking of Archean, Proterozoic, and modern-Earth
detectability for an HWO-like telescope is beyond the scope of this
paper, but Eq. (\ref{eq:ratio_final}) clarifies the terms of that comparison: first
detection depends on the product of abundance and detectability, not on
longevity alone.

The Earth through time cases discussed above are illustrative, not
exhaustive. The first HWO biosignature need not correspond to any
spectral state previously realized on Earth. A rocky inhabited planet
with a non Earth-like atmospheric composition, surface environment, or
biogenic flux could in principle be even more detectable than the
Earth through-time analogs considered here. The central point is
therefore broader than the Earth-through-time comparison alone: HWO is
most likely to detect first the inhabited rocky planet whose
biosignature spectrum is easiest to observe, not necessarily the one
whose atmospheric state is most familiar or most common.



\section{Why the First Biosignature May Not Be Representative}

The HWO case differs from the JWST case in an important way, but it is
still governed by the same general lesson illustrated by earlier
astronomical first detections. The first pulsar planets, the first hot
Jupiter, and the first detection of exoplanet atmospheres were all
selection favored objects rather than representative members of their
parent populations.  In other words, HWO will probably detect first the inhabited rocky planet that is easiest to see, but not the one that is most common. From Equations~(\ref{eq:contrast_signal}) and
(\ref{eq:contrast_feature}), one can see that a very strong HWO signal
could in principle come from a highly reflective sub-Neptune, whose
larger radius and elevated albedo, for example from water clouds or
hazes, would increase its reflected light contrast. In that sense,
the direct imaging case may not be entirely different from the
transmission spectroscopy case, where large radius, atmosphere rich
planets are likewise favored; K2-18b is an illustrative example.

 It should be noted that the biosignature molecules most often discussed for sub-Neptune Hycean worlds, such as DMS and DMDS, have their strongest spectral features at 6–11$\mu$m \citep{tsai2024}, well outside HWO’s 0.2–2.0$\mu$m bandpass. While a larger planetary radius does increase the reflected light contrast $C_{p}$ (Eq. (\ref{eq:contrast_signal})), the corresponding biosignature feature depth $A_{bio}$
at HWO wavelengths may be much smaller and is not yet well characterized. The contrast advantage of large radius sub-Neptunes therefore does not automatically translate into a comparable detectability advantage for Hycean-world biosignatures.

For JWST,
the most favorable biosignature targets are likely to be planets that
are already extreme in their basic observational properties, such as
sub-Neptunes or H$_2$-rich worlds orbiting small stars. For HWO, the
problem is subtler. HWO is designed to target rocky planets around
nearby Sun-like (FG) and K-type stars, but those planets may span a broad
range of atmospheric states. The character of the first HWO
biosignature detection therefore depends not only on instrumental
performance and target-star properties, but also on which
Earth-through-time atmospheric state happens to produce the most
detectable spectral signature in the accessible survey sample.

The key point is that the first HWO biosignature should not be assumed
to be representative of the broader population of inhabited rocky
planets. If the first HWO biosignature resembles modern Earth
(\Ot{}+\Oz{}+\HzO{}), it would be tempting to interpret that result as
evidence that modern-Earth-like biospheres are the dominant inhabited
state of rocky planets. But that inference may not be secure. Modern
oxic atmospheres may instead be over-represented among early detections
simply because their optical and near-infrared spectral features are
especially prominent within HWO's wavelength range. Conversely, if the
first HWO biosignature resembles an Archean- or Proterozoic-like state
(for example \CH{}+\CO{} disequilibrium or low-\Ot{} with detectable
\Oz{}), that would likewise not imply that such biospheres dominate the
inhabited-planet population. A single nearby planet with favorable
distance, cloud properties, and atmospheric composition could be the
most detectable member of the accessible sample while remaining
unrepresentative of inhabited rocky planets as a whole.

Indeed, because HWO is being designed in part to enable the detection
of O$_2$ and O$_3$ bearing atmospheres, the mission architecture may
itself reinforce a selection effect in favor of oxic biosignatures,
even if such states are not the dominant inhabited outcome.

The Earth through time cases discussed above are illustrative, not
exhaustive. The first HWO biosignature need not correspond to any previous 
climate state on Earth. A rocky inhabited planet
with a non Earth-like atmospheric composition, surface environment, or
biogenic flux could in principle be even more detectable than the
Earth through time analogs considered here. The central point is
therefore broader than the Earth through time comparison alone: HWO is
most likely to detect first the inhabited rocky planet whose
 spectrum is easiest to observe, not necessarily the one
whose atmospheric state is most familiar or most common.

In this sense, the first HWO biosignature is likely to be the most
detectable Earth-like biosphere in the local survey volume, rather than
the most common one. This creates an important interpretive challenge:
the fact that HWO is designed to search for Earth-like life around
nearby Sun-like stars does not imply that its first biosignature
detection will be typical of Earth-like biospheres in general. The same
logic applies to the type of biosignature detected. Among inhabited
rocky planets accessible to HWO, those with the largest spectral
feature depths will be favored in early detections, regardless of
whether those biosignature strengths are typical. A planet with an
unusually strong Archean-like \CH{} signature, for example, could be
detected before a more representative modern-Earth analog at the same
distance.

A similar point applies to atmospheric technosignatures. For example,
NO$_2$ has been proposed as a potentially detectable industrial
pollutant in reflected light \citep{kopparapu2021}, and if present at sufficiently high
abundance could in principle be easier to detect than some weaker
biosignature states. We do not pursue that comparison quantitatively
here, but it illustrates the broader point that first detections are
expected to favor spectrally prominent atmospheric states, not
necessarily familiar or representative ones.

 A related complication is that the atmospheric states favored in early
detections may also be among the most difficult to interpret. If the
first biosignatures arise from spectrally extreme or unfamiliar worlds,
then the same properties that make them easier to detect may also
increase the difficulty of distinguishing biological, abiotic, and
possibly technological explanations. In that sense, first detection
bias and false positive risk may be linked: the first atmospheric
signatures to attract attention may be the least representative and the
most controversial. This argues for a biosignature framework that
remains as agnostic as possible about familiar Earth analogies while
requiring robust contextual vetting. Furthermore, a framework for evaluating biosignature claims would also help in validitating initial discoveries \citep{meadows2022}.

\subsection{The Non-Detection Inference Problem}
\label{subsec:nondetect}
 
The ``first detection'' framework proposed here also clarifies how
non detections should be interpreted. A JWST survey that fails to
detect biosignatures in the atmospheres of the most favorable
M-dwarf habitable zone targets does not imply that life is rare. It
implies only that the most detectable cases in the local neighborhood
do not host biosignatures above the detection threshold, which is a
much narrower statement. Similarly, if HWO surveys habitable zone
rocky planets around nearby Sun-like stars and finds none with
biosignatures, this constrains the frequency of life among the most
detectable HWO targets, not among rocky planets as a whole.

The same caution applies to atmospheric non detections themselves.
JWST may show that several temperate terrestrial planets around
M dwarfs have little or no detectable atmosphere, but that would not
mean that all such planets are bare. Following the same logic as the
historical examples discussed in the Introduction, the first planets
for which atmospheric loss is clearly established may simply be the
easiest cases in which that outcome can be observed, rather than
representative members of the broader population. In that sense, an
early sample dominated by  planets with no atmosphere would not necessarily
show that such planets are typical; it may instead reflect the same
first detection bias that has shaped many earlier discoveries in
astronomy. Planets with atmospheres around M dwarfs could still be
common even if the first well constrained cases turn out to be bare.

This point is especially relevant because some of the strongest current
JWST constraints on rocky exoplanet atmospheres come from hot
terrestrial planets observed in eclipse, which are also among the
planets most likely to have undergone substantial atmospheric loss. In
that sense, early atmospheric non detections may be sampling a
particularly unfavorable subset of rocky planets rather than the class
as a whole.

\section{Conclusions}
\label{sec:conclusions}

We proposed how observational selection effects shape the
character of the first remotely detectable biosignatures for the two
leading near future search strategies: transmission spectroscopy with
JWST and reflected light direct imaging with HWO.


  Transmission spectroscopy and reflected-light direct imaging
        favor different regions of parameter space. 
K2-18b illustrates the kind of target that current
        transmission spectroscopy capabilities favor. Whether or not it
        ultimately proves habitable, it lies near the top of the
        present detectability hierarchy and therefore serves as a
        useful example of the kind of system likely to generate one of
        the first claimed biosignature detections.

  For HWO, the  problem is nuanced. Even within a
        survey restricted to rocky planets around nearby Sun like and
        K-type stars, the first biosignature detected may still be
        unrepresentative of inhabited rocky planets as a whole. It is
        likely to be the most detectable atmospheric state in the
        accessible sample, not necessarily the most common one.

  This point applies directly to Earth-through-time scenarios. If
        the first HWO biosignature resembles modern Earth, that does
        not by itself imply that modern Earth like biospheres are common among rocky planets. If it instead resembles an Archean- or
        Proterozoic-like atmosphere, that equally does not show that
        such biospheres are typical. In either case, early detections
        should be interpreted in the context of selection effects
        before being used to infer the prevalence of life.

  The same caution applies to non detections. Early JWST atmospheric non-detections of rocky planets around M dwarfs would not show
        that such planets generally lack atmospheres, only that the
        first observed cases may be biased toward the easiest
        systems in which atmospheric loss can be established. Similarly, a  null result  from
        JWST observations of favorable M-dwarf targets, or from an HWO
        survey of nearby rocky planets, constrains the most detectable
        cases first, not the full inhabited-planet population. 
        

Taken together, these results suggest that the first remote
biosignature detection should be interpreted not as a typical example
of life beyond Earth, but as the product of a particular observing
strategy acting on a diverse planetary population. The first inhabited
world we find may therefore tell us as much about the selection effects
of our instruments as it does about life in the universe.

There is also an encouraging implication of this argument. Even if the first detection turns out to be an outlier, it would still suggest that a wider and potentially very diverse population of habitable environments awaits discovery. In that sense, the first detection would not be the endpoint of the search, but the first glimpse of a broader community of habitable worlds.

\begin{acknowledgments}

R.K acknowledges support from the GSFC Sellers Exoplanet Environments Collaboration (SEEC), which is funded by the NASA
Planetary Science Division's Internal Scientist Funding Model.  This material is also based upon work performed as part of the CHAMPs (Consortium on Habitability and Atmospheres of M-dwarf Planets) team,
supported by the National Aeronautics and Space Administration (NASA) under Grant No. 80NSSC23K1399
issued through the Interdisciplinary Consortia for Astrobiology Research (ICAR) program. 

The author is grateful to the anonymous reviewer whose thoughtful comments substantially improved the clarity, quality, and readability of the manuscript.

The author thanks Prabal Saxena for insisting on submitting this work. The author is grateful for valuable input from Jacob Haqq-Misra, Jacob Lustig-Yaeger, Kevin Stevenson, Eddie Schwieterman, and Noah Tuchow who reviewed the original manuscript and/or provided comments.

The author thanks the organizers and participants of the Exploring Exoplanets school at the International Centre for Theoretical Sciences (ICTS), Bangalore, for their hospitality and for discussions that contributed to this work.

The author also wishes to thank colleagues from the `Astronomers’ Facebook community who responded to my query of March 6, 2020, regarding outlier events in astronomical discoveries. That discussion pointed to many additional examples across several subfields of astronomy, far too numerous to list in this manuscript. Interested readers may locate the original post and the associated discussion by searching for the author’s name within the community group.

Data files used to generate Figure 2 are available with the manuscript.

{\it Software:} 
The author acknowledges the use of LLMs  (ChatGPT and Claude)  as a tool to assist with grammar, sentence construction, and clarity of expression. All scientific arguments, interpretations, and references were developed and verified by the author.
\end{acknowledgments}

\bibliography{sample631}{}
\bibliographystyle{aasjournal}

\end{document}